# Understanding the two-step nucleation of iron at Earth's inner core conditions: a comparative molecular dynamics study

Chen Gao[1], Kai-Ming Ho[2], Renata M. Wentzcovitch[3-5], and Yang Sun[1*]

[1]*Department of Physics, Xiamen University, Xiamen 361005, China*
[2]*Department of Physics, Iowa State University, Ames, IA 50011, USA*
[3]*Department of Applied Physics and Applied Mathematics, Columbia University, New York, NY 10027, USA*
[4]*Department of Earth and Environmental Sciences, Columbia University, New York, NY, 10027, USA*
[5]*Lamont–Doherty Earth Observatory, Columbia University, Palisades, NY, 10964, USA*

Metastable phases can lead to multistep nucleation processes, influencing the liquid-to-solid transition in various systems. In this study, we investigate the homogeneous nucleation of iron's crystalline phases under Earth's inner core conditions, employing two previously developed interatomic potentials. We compare the thermodynamic and kinetic properties of iron relevant to the nucleation as predicted by these potentials. While the potentials differ in their predictions of melting temperature by a few hundred Kelvin, they show a consistent description of the relative Gibbs free energy between solid and liquid phases with respect to the undercooling. Both potentials also predict that the metastable bcc phase exhibits a significantly higher nucleation rate than the hcp phase over a wide range of undercooling temperatures below the melting point. This substantially lowers the undercooling thresholds required for the initial nucleation of Earth's inner core. The results validate the commonality of the two-step nucleation mechanism of iron under Earth's inner core conditions for two different potentials, providing a foundation for future studies about the influence of other elements on the nucleation of Earth's core.

## I. INTRODUCTION

Crystal nucleation from a liquid is a common phenomenon that impacts various fields, from materials science to biophysics [1,2]. Classical nucleation theory (CNT) [3] is widely used to understand this process, which is characterized by the competition between the driving force of the bulk free energy and the energy penalty associated with solid-liquid interface (SLI) formation. The simplest scenario in CNT assumes that a crystal nucleus forms due to structural fluctuations in the liquid. Once the nucleus overcomes the energy barrier to reach the critical size in the undercooled liquid, it will continue to grow until the entire liquid solidifies or it encounters another growing crystal grain. CNT describes the nucleation without any competing phases, which is oversimplified. The situation becomes more complex when multiple crystal phases can nucleate within the liquid phase. The thermodynamically stable phase is not necessarily the first to nucleate [4–12]. Metastable phases can nucleate before the stable phase emerges and later transform into the stable phase, effectively lowering the nucleation barrier. Such multistep nucleation processes have been observed in synthetic systems, biominerals, and natural mineralization patterns [13,14].

Recently, the nucleation of melts in Earth's core has attracted significant interest because of its relation to the age and formation process of Earth's inner core (IC) [15]. Applying classical nucleation theory to Fe under IC conditions suggests that an exceptionally large undercooling, up to 1000 K, is required to overcome the nucleation barrier. Given the slow cooling rate throughout Earth's core history [16], reaching such a large undercooling within the plausible age of the core is impossible. Therefore, this prediction fails to explain the current existence of Earth's IC, giving rise to the so-called "inner core nucleation paradox" [17]. However, making predictions based solely on the CNT formula is non-trivial, as any specific CNT calculation relies on input parameters such as bulk free energy and solid-liquid interface free energy, which are often difficult to obtain. For example, obtaining the thermodynamic data or SLI free energy becomes very challenging if a metastable phase exists only for a relatively short period. The anisotropy of the SLI free energy and mobility adds further complexity, as one must account for the shape of the nucleus [18,19]. Without this information, CNT cannot accurately predict the outcome of phase competition.

The coupling of CNT with molecular dynamics (MD) simulations helps circumvent some of these difficulties. By applying the seeding method to simulate nucleation of the stable hcp phase, Davies *et al.* demonstrated that the undercooling required for Fe nucleation in the core can be reduced to approximately 800 K [20]. Wilson *et al.* extrapolated the radius of subnuclei and found a similar estimate for the required undercooling for IC nucleation [21]. While this value is lower than that obtained from direct calculations with CNT formulas by Huguet *et al.* [17], it is still far outside core's possible undercooling range of 100~200 K, which was estimated by the core's cooling rate [16] and age [20]. Other elements, such as carbon [22] and nickel [23], have been found to accelerate Fe's nucleation, but their effects highly depend on their concentrations in the core, which remain poorly constrained [24].

[*]Email: yangsun@xmu.edu.cn

Sun *et al.* examined the nucleation of metastable phases under core conditions [25]. They found that the metastable bcc phase has a much higher nucleation rate than the stable hcp phase under IC conditions. If the core nucleates the bcc phase first, the required undercooling can be further reduced to approximately 500 K [25]. This two-step nucleation mechanism provides a new scenario for the IC formation process [26]. However, the nucleation of the metastable bcc phase was not reported in the simulations performed by Wilson *et al.* using brute-force MD methods [21,22].

If the IC grew from an undercooled liquid, its initial growth may have been more rapid than its current rate, leading to fluid entrapment in the deep early IC [27]. These fluids can undergo two-stage crystallization, resulting in textures distinct from those of the upper IC and enhancing seismic scattering in the deep IC [28]. The two-step nucleation can also create a coexistence region of different crystalline phases near the inner core boundary (ICB), potentially leading to complex ICB structures and seismic anomalies [29]. Given these significant geophysical implications, it is crucial to assess the generality of two-step nucleation for future studies of core crystallization involving complex Fe alloys. One of the main differences between the work of Sun et al. [25] and Wilson *et al.* [21] lies in their use of different interatomic potentials for Fe [25,30]. These interatomic potentials are crucial for performing MD simulations at large length scales and long timescales, which are required to observe nucleation. However, the outcomes of such simulations can be highly sensitive to the quality of these potentials [31,32]. Additionally, the two studies employed different methods to identify the crystalline order in the as-formed nuclei. In this work, we aim to clarify the similarities and differences between the two interatomic Fe potentials used in these studies to describe Fe nucleation under Earth's core conditions. We employ the same methods to simulate the nucleation of both the stable hcp phase and the metastable bcc phase, analyzing the results with a consistent methodology.

This paper is organized as follows. In Section II, we outline the methods used for free energy and nucleation rate calculations, along with the details of the MD simulations. In Section III, we compare the results obtained using the two interatomic potentials, including melting temperature, bulk free energy, solid-liquid interface energy, and nucleation rate. In Section IV, we discuss the similarities and differences in the properties of Fe simulated by the two potentials and their impact on predicting Fe's nucleation rate under IC conditions. Finally, we present our conclusions in Section V.

## II. METHODS

### A. Melting temperature and solid-liquid free energy difference

The melting temperatures $T_m$ were computed using the solid-liquid coexistence approach [33]. Based on $T_m$, the Gibbs-Helmholtz equation was applied to calculate the bulk free energy difference between solid and liquid phases ($\Delta\mu = \mu^{\text{solid}} - \mu^{\text{liquid}}$) as

$$\frac{\Delta\mu(T)}{T} - \frac{\Delta\mu(T_m)}{T_m} = -\int_{T_m}^{T} \frac{\Delta H(T)}{T^2} dT, \quad (1)$$

where $\Delta H$ is the enthalpy difference between solid and liquid ($\Delta H = H^{\text{solid}} - H^{\text{liquid}}$). With $\Delta\mu(T_m) = 0$, $\Delta\mu(T)$ can be computed as

$$\Delta\mu(T) = -T \int_{T_m}^{T} \frac{\Delta H(T)}{T^2} dT. \quad (2)$$

### B. Classical Nucleation Theory

In the CNT, the Gibbs free energy change ($\Delta G$) during the nucleation process is determined by the driving force from the bulk free energy difference ($\Delta\mu$) and the energy penalty associated with the formation of the solid-liquid interface. This relationship is described as

$$\Delta G = N\Delta\mu + A\gamma, \quad (3)$$

where $N$ denotes the number of atoms in the nucleus, and $\gamma$ is the SLI free energy per unit area. $A$ represents the interface area, which can be evaluated as $A = s(N/\rho_c)^{2/3}$, where $\rho_c$ is the crystal density and $s$ is the shape factor. The competition between the bulk and interface terms leads to a nucleation barrier $\Delta G^*$ when the nucleus reaches the critical size $N^*$, i.e., $\frac{\partial \Delta G(N^*)}{\partial N} = 0$, and

$$\Delta G^* = \frac{4s^3\gamma^3}{27|\Delta\mu|^2\rho_c^2}. \quad (4)$$

In CNT, it is typically assumed that the nucleus has a spherical shape ($s_{CNT} \equiv \sqrt[3]{36\pi}$) in order to compute $\Delta G^*$ with known values of $\gamma$ and $\Delta\mu$. However, this assumption can be relaxed by computing the shape factor $s$ from simulations. If the critical nucleus can be directly obtained from the simulations, the four key quantities ($\rho_c$, $\Delta\mu$, $N^*$, and $s$) can be determined, which allows for the calculation of the SLI free energy $\gamma$ as

$$\gamma = \frac{3}{2s}|\Delta\mu|\rho_c^{2/3}N^{*\frac{1}{3}}. \quad (5)$$

Combining Eqn. (4) and (5) yield a simplified expression for the nucleation barrier $\Delta G^* = \frac{1}{2}|\Delta\mu|N^*$. Then, the nucleation rate $J$ can be obtained by

$$J = \kappa \exp\left(-\frac{\Delta G^*}{k_B T}\right), \quad (6)$$

where $\kappa$ is a kinetic prefactor and $k_B$ is the Boltzmann constant. The perfector $\kappa$ can be derived from the steady-state model [1] as

$$\kappa = \rho_L f^+ \sqrt{\frac{|\Delta\mu|}{6\pi k_B T N^*}}, \quad (7)$$

where $f^+$ is the attachment rate of a single atom to the critical nucleus and $\rho_L$ is the liquid density.

### C. Persistent embryo method

The persistent embryo method (PEM) [34] was employed to accelerate the nucleation simulation and determine both the critical nucleus size $N^*$ and the shape factor $s$. The PEM is based on the fundamental CNT concept that homogeneous nucleation occurs via the formation of a critical nucleus in the undercooled liquid. The PEM allows efficient sampling of the nucleation process by preventing a small crystal embryo (with $N_0$ atoms, much smaller than the critical nucleus) from melting through the use of the external spring forces. This eliminates prolonged periods of ineffective simulation where the system is far from forming a critical nucleus. As the embryo grows, the harmonic potential is gradually weakened and completely removed when the cluster size reaches a sub-critical threshold, $N_{sc}$, which is smaller than $N^*$. During the simulation, the harmonic potential applies only to the original $N_0 (< N_{sc})$ embryo atoms. The spring constant of the harmonic potential decreases with increasing nucleus size, given by $k(N) = k_0 \frac{N_{sc}-N}{N_{sc}}$ if $N < N_{sc}$ and $k(N) = 0$ otherwise. This ensures the system remains unbiased at the critical point, allowing for a reliable determination of the critical nucleus size. If the nucleus melts below $N_{sc}$ ($< N^*$), the harmonic potential is gradually reinforced, preventing the complete melting of the embryo. When the nucleus reaches the critical size, it has an equal probability of melting or growing further, resulting in fluctuations around $N^*$. As a result, the $N(t)$ curve tends to display a plateau during critical fluctuations, providing a unique signal to detect the appearance of the critical nucleus. The critical nucleus size is calculated by averaging $N(t)$ over such a plateau. Multiple plateaus can be observed before a nucleus grows. For statistical analysis of the critical nucleus's size and shape, four plateaus with long fluctuation time are selected, as shown in Supplemental Material Fig. S1 [35].

Based on the nucleus obtained via PEM, the attachment rate can be computed as the effective diffusion constant for the size change of the critical nucleus, given by $f^+ = \frac{\langle|\Delta N^*(t)|^2\rangle}{2t}$ [36]. The iso-configurational ensemble simulation [37] was employed to measure $f^+$ following the PEM-MD simulations. In the iso-configurational ensemble simulation, 30 independent MD runs were launched, each initiated from the same atomic configuration with atomic momenta randomly assigned based on the Maxwell-Boltzmann distribution. The initial configurations nuclei were collected from the plateaus in Supplementary Material Fig. S1 [35]. If the critical nuclei melt in half of the iso-configurational ensemble simulations and grow in the other half statistically, the determination of the critical nucleus size can be validated. The results of iso-configurational ensemble simulations are shown in Supplemental Material Fig. S3 [35].

### D. Brute-force simulation

Since nuclei of different phases exhibit different nucleation rates, the liquid system statistically favors the formation of the solid phase with the higher nucleation rate. Brute-force simulations were conducted to determine which phase nucleates first at very large undercooling. In the brute-force simulation, a liquid is equilibrated at the melting temperature for 100 *ps* and quenched to a temperature below the melting point, followed by a 10 *ns* NPT simulation. During this simulation, we monitored changes in the size and structure of spontaneously formed nuclei over time.

### E. Molecular dynamics simulation

MD simulations were performed using the Large-scale Atomic/Molecular Massively Parallel Simulator (LAMMPS). Two semi-empirical potentials developed for high-pressure Fe by Alfè et al. [20,30] and Sun et al. [25] were employed in the simulations, hereafter referred to as PotA and PotM, respectively ("M" indicates the main developer, M. I. Mendelev, for the potential in Ref. [25]). To maintain consistency with prior investigations [20,25], the simulations were performed at a pressure of 323 GPa, approximating conditions near the inner core boundary. The system size is 16,000 atoms for solid-liquid coexistence simulations and approximately 31,250 atoms for nucleation simulations.

The polyhedral template matching (PTM) [38] method was employed to analyze the local structural environment of the particles to classify the solid-like atoms by computing the similarities between atomic clusters and the perfect template clusters of bcc and hcp crystals. The classification threshold is determined using the equal mislabeling method, as described by Espinosa *et al.* [39]. We note that a small fraction of fcc clusters can occasionally form at the solid-liquid interface of the nucleus during PEM simulations, which are considered stacking faults of the hcp phase.

## III. RESULTS

### A. Bulk free energy

We first compare the solid and liquid free energy difference between the two potentials, which provides the driving force of nucleation. Solid-liquid coexistence simulations are performed to obtain the melting points for the hcp and bcc phases using the two potentials, as summarized in Table 1. We find the melting point of hcp obtained from the current simulation is consistent with previous work for PotA [20] and PotM [25]. Both potentials show that hcp has a higher melting temperature than bcc, suggesting that the hcp phase is stable for pure Fe at 323 GPa, while the bcc phase is metastable. The differences in melting points between bcc and hcp are almost the same for the two potentials, approximately 70 K. However, the melting temperatures from PotM are approximately 300 K lower than those from PotA. This discrepancy is primarily due to the fact that PotM was fitted to an experimental melting data [40] which are lower than the *ab initio* results.

Table 1. The melting temperature of the bcc and hcp phases computed with PotA and PotM potentials at 323 GPa.

| Potential | hcp | bcc |
|---|---|---|
| PotA | 6215 K | 6130 K |
| PotM | 5858 K | 5793 K |

The free energy difference $\Delta\mu$ is computed using the Gibbs-Helmholtz equation (Eqn. (2)). Figure 1(a) shows the temperature-dependent $\Delta\mu$ for the two potentials, illustrating the phase competition among hcp, bcc and liquid. Both potentials indicate that bcc has higher free energy than hcp over a large temperature range below $T_{\mathrm{m}}$. Since the two potentials have different $T_{\mathrm{m}}$, it is more reasonable to compare the free energy difference at the same undercooling ($\Delta T = T - T_m$). Figure 1(b) shows $\Delta\mu$ as a function of the relative undercooling temperature with respect to the melting temperature of the hcp phase. At 200 K undercooling, the free energy difference between the bcc and hcp phase is 7 meV/atom for PotA and 9 meV for PotM. When the undercooling increases to 1000K, the free energy difference between the bcc and hcp phase is 13 meV/atom for PotA but 27 meV/atom for PotM. Therefore, at small undercooling, the two potentials provide similar estimates of the relative stability between the hcp and bcc phases. At larger undercooling, the difference in relative stability becomes larger in PotM than in PotA.

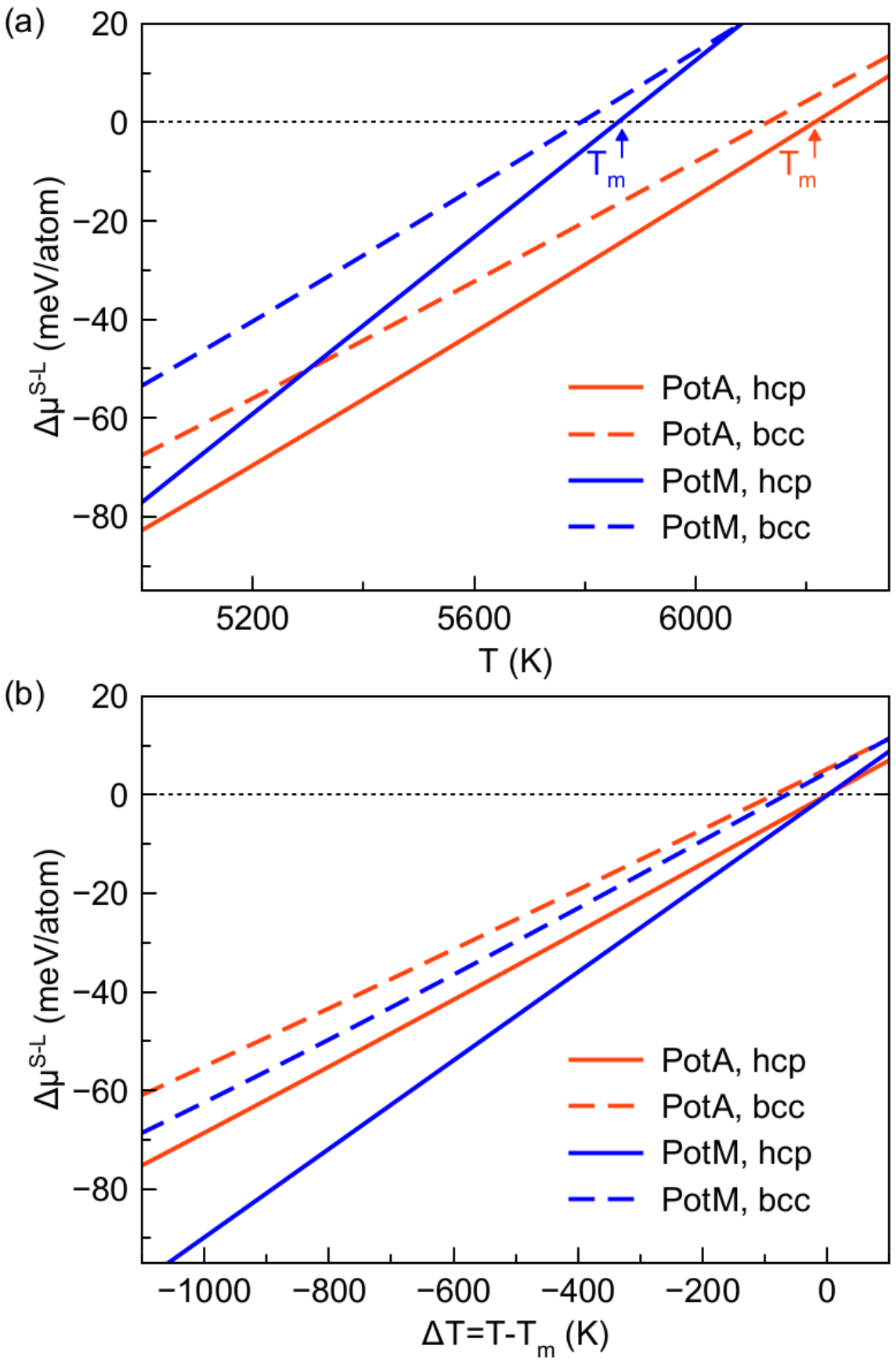

FIG. 1. The Gibbs free energy difference between solid and liquid as a function of (a) temperature and (b) undercooling relative to the melting temperature of hcp for PotA and PotM. The black dotted line indicates $\Delta\mu = 0$.

### B. Nucleation simulation

PEM simulations were previously performed for PotM [25]. Here, we perform PEM simulations for PotA and analyze the differences between the two potentials. The trajectories used to identify the critical nucleus from PEM simulations are provided in the Supplemental Material Fig. S1 [35]. Figure 2(a) shows the critical nucleus sizes of the bcc and hcp phases at 323 GPa for both potentials. For both PotM and PotA, hcp always has a larger critical nucleus than bcc. The data are also plotted with respect to the undercooling in Fig. 2(b). The comparison between PotA and PotM suggests that PotA leads to a larger critical nucleus than PotM for both hcp and bcc phases at the same undercooling. Figure 2(b) also compares the critical nucleus size data with the ones from previous work using PotA [20,21]. The critical nucleus sizes from Davies *et al.* [20] using the seeding method are close to the current PEM results. Wilson *et al.* [21] estimated the critical nucleus size by measuring the radius and formation rate of the sub-critical nucleus. Because the rate of forming a nucleus exponentially depends on the nucleus size, this method can be extremely computationally expensive to collect sufficient statistical data to fit a reasonable free energy distribution. In Fig. 2(b), the data from Wilson *et al.* [21] show a significantly large error bar and are systematically much higher than our PEM results and the data from Davies *et al.* [20].

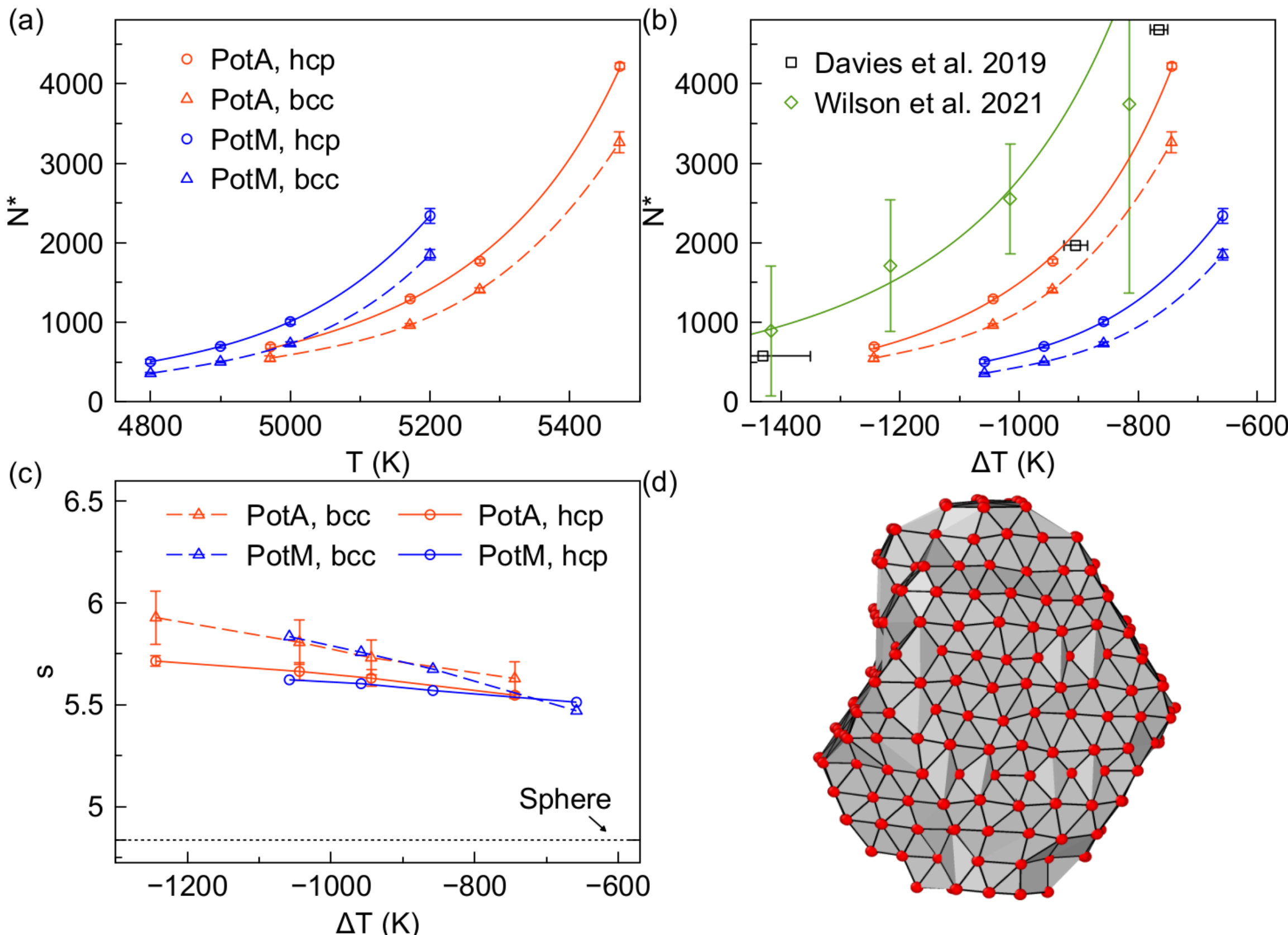

FIG. 2. The critical nucleus size as a function of (a) temperature and (b) undercooling relative to the melting temperature of hcp for PotA and PotM at 323 GPa. Black and Green markers are previous data computed with PotA by Davies *et al.* [20] and Wilson *et al.* [21]. The green solid line represents the fits from Ref. [21], and the blue and red lines are obtained by polynomial fitting. (c) The shape factor of the critical nucleus as a function of undercooling for PotA and PotM at 323 GPa. The black dotted line indicates the shape factor of a perfect sphere which equals to $\sqrt[3]{36\pi}$. (d) An irregular bcc nucleus configuration, which is not spherical, from the PEM simulations of PotM at a temperature of 4800 K. The error bars are calculated based on the standard deviations of four independent calculations.

Davies *et al.* [20] and Wilson *et al.* [21] both assume a spherical shape for the nucleus. Since PEM simulations do not constrain the nucleus, we can measure the shape of the critical nucleus spontaneously formed in PEM simulations (see details in Supplemental Material Text S1 [35]). The shape factors were computed as $s = A/V^{2/3}$ [41], where $A$ is the surface area and $V$ is the volume of the nucleus polyhedron, constructed by the geometric surface reconstruction method [42] integrated in the OVITO software package [43]. Examining Fig. 2(c) shows that the shape factors of both hcp and bcc critical nuclei deviate from that of a perfect sphere. The shape factors computed with PotA and PotM are similar and exhibit similar temperature dependences. At higher temperatures, the shape factor of the critical nucleus is closer to that of a sphere. As the temperature decreases, the shape factor becomes larger, indicating that the nucleus is more anisotropic as its size decreases. Figure 2(d) shows an example of a critical nucleus with a shape quite different from a sphere and facets at the interface.

The SLI free energy $\gamma$ can be calculated with the measured shape factor and the critical size by Eqn. (5). Figure 3(a) and (b) show the $\gamma$ of hcp and bcc phases for both PotA and PotM. In both systems, the SLI free energy shows a nearly linear temperature dependence. Such linear temperature dependence is similar to the previously measured $\gamma$ for Al and Ni [41]. The $\gamma$ data can be extrapolated to higher temperatures and compared with the previous results obtained using the superheating-supercooling hysteresis method at the same pressure [44]. Even though $\gamma$ from [44] was determined using a completely different method and interatomic potential, it deviates by only 4% from the extrapolated data of both PotA and PotM. Figure 3(a) shows that the SLI free energy of hcp is larger than that of bcc for both PotA and PotM. In Fig. 3(b), at the same undercooling, PotA has larger SLI free energy than PotM for the bcc phase. For the hcp phase, the $\gamma$ data of PotA and PotM are similar and overlap at ~1000 K undercooling.

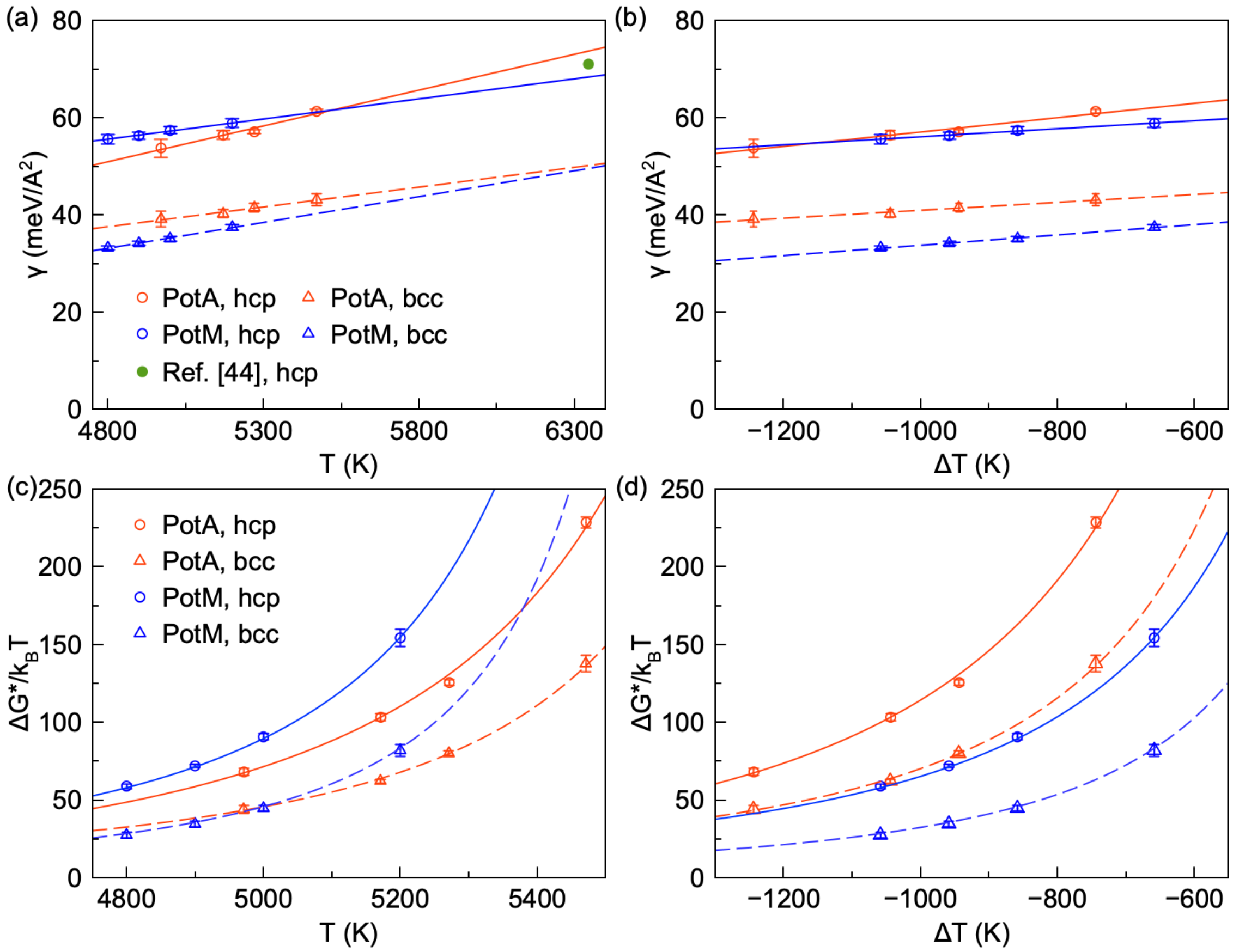

FIG. 3. The SLI free energy of hcp and bcc phases as a function of (a) temperature and (b) undercooling relative to the melting temperature of hcp for PotA and PotM at 323 GPa. (c) The free energy barrier as a function of temperature. (d) The free energy barrier as a function of undercooling.

Combining the temperature-dependent chemical potential, SLI free energy, and shape factor, the nucleation barrier can be computed via Eqn. (4). Figure 3(c) and (d) show $\Delta G^*$ as functions of temperature and undercooling, respectively, for both the hcp and bcc phases. For both potentials, the nucleation barrier of hcp is always higher than that of bcc, suggesting that bcc is always easier to nucleate than hcp at this range of temperatures. At the same undercooling, the nucleation barrier of PotA is always higher than that of PotM for both hcp and bcc phases, as shown in Fig. 3(d). Such differences between the two potentials are caused by the complex competition in the chemical potential and SLI free energy. For hcp, the nucleation driving force (bulk free energy difference) of PotA is smaller than that of PotM, as shown in Fig. 1(b), while its SLI free energy is approximately the same with PotM at the same undercooling in Fig. 3(b). For the bcc phase, the smaller driving force and stronger energy penalty of SLI in PotA lead to larger nucleation barriers than those in PotM.

Based on the critical nuclei from PEM simulations, we compute the attachment rate, $f^+$, using the iso-configurational MD simulations as the effective diffusion constant for the change in the critical nucleus size [34,36,45]. The MD results are shown in

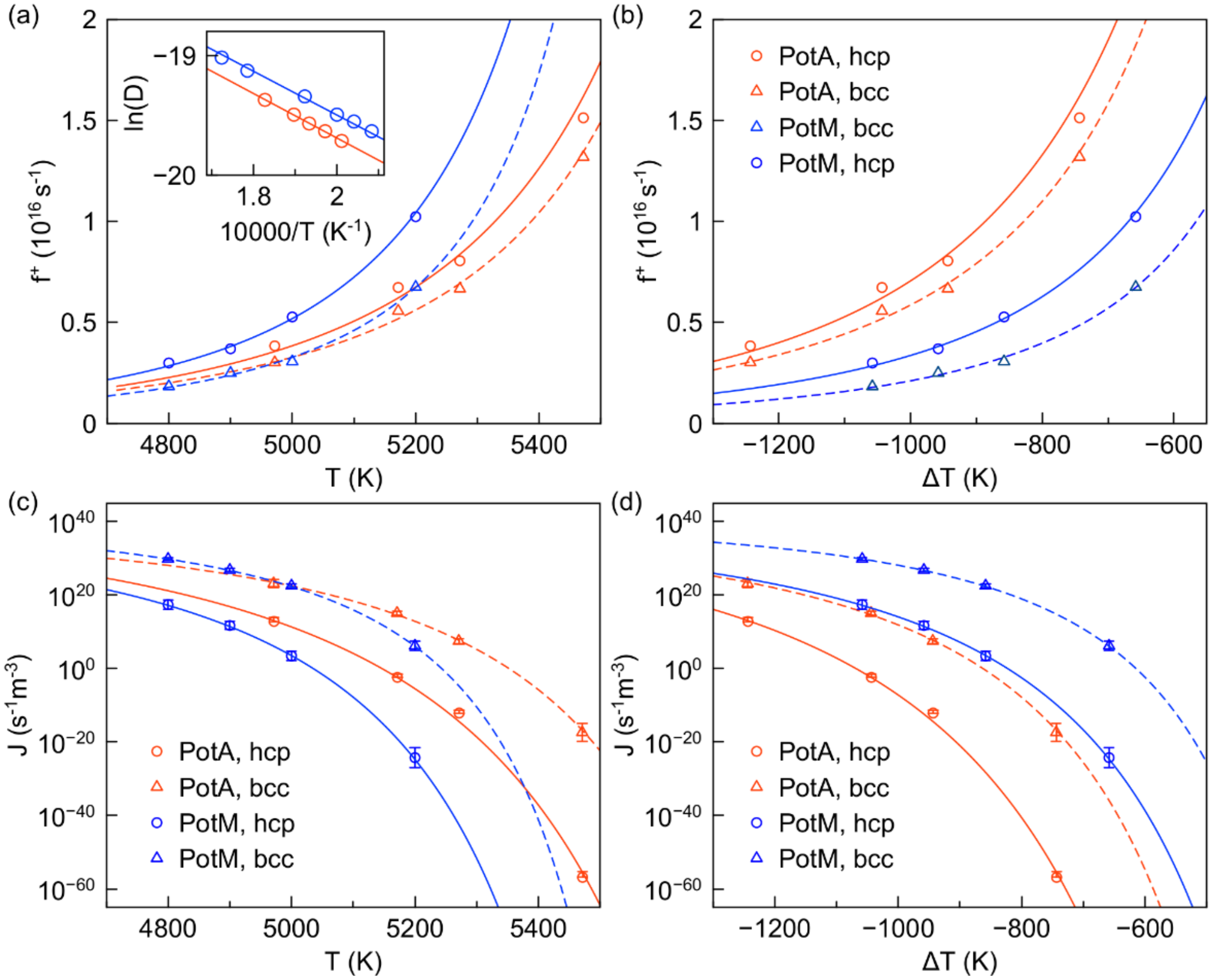

FIG. 4. The attachment rate as a function of (a) temperature and (b) undercooling relative to the melting temperature of hcp at 323 GPa. The solid and dashed lines are fitted by Eqn. (8). The inset in (a) shows the liquid diffusivity as a function of temperature, with the blue and red circles representing the results of PotM and PotA, respectively. The nucleation rate as a function of (c) temperature and (d) undercooling relative to the melting temperature of hcp at 323 GPa. The lines are computed with Eqn. (6).

Supplemental Material Fig. S3 [35], for both the hcp and bcc phases with PotA and PotM. To interpolate the data with temperature dependence, we employ the classical kinetic model of atom attachment [1] where $f^+$ is proportional to the liquid diffusivity $D$ and the nucleus surface area. With the shape factor $s$, the expression for $f^+$ can be written as

$$f^+ = sN^{*2/3}\frac{6D}{\lambda^2}, \quad (8)$$

where $\lambda$ is the atomic jump distance during the attachment, which can be determined based on the measured $f^+$. For the undercooled Fe liquids considered here, the temperature dependence of the bulk diffusion coefficient can be well fitted to the Arrhenius relationship [46], as shown in the inset of Fig. 4(a). With all the parameters in Eqn. (8) available, the attachment rate is interpolated over a wide temperature range. The attachment rate of the hcp nucleus is typically faster than that of the bcc nucleus at the same temperature. The difference in $f^+$ between bcc and hcp decreases with undercooling. These behaviors are consistently described by both PotA and PotM potentials. At the same undercooling, the attachment rate of PotA is slightly faster than that of PotM, while the difference between the two potentials gradually narrows as the temperature decreases.

As all the parameters and their temperature dependencies related to Eq. (6) are determined, the temperature-dependent nucleation rate is computed for both the hcp and bcc phases with PotM and PotA. As shown in Fig. 4(c) and (d), the bcc phase shows a systematically faster nucleation rate than the hcp phase. At high temperatures or small undercooling, the difference in nucleation rates between the hcp and bcc phases is significant, while it becomes much smaller at lower temperatures and larger undercooling. This behavior is consistent for both PotM and PotA potentials. The nucleation rate has a systematic

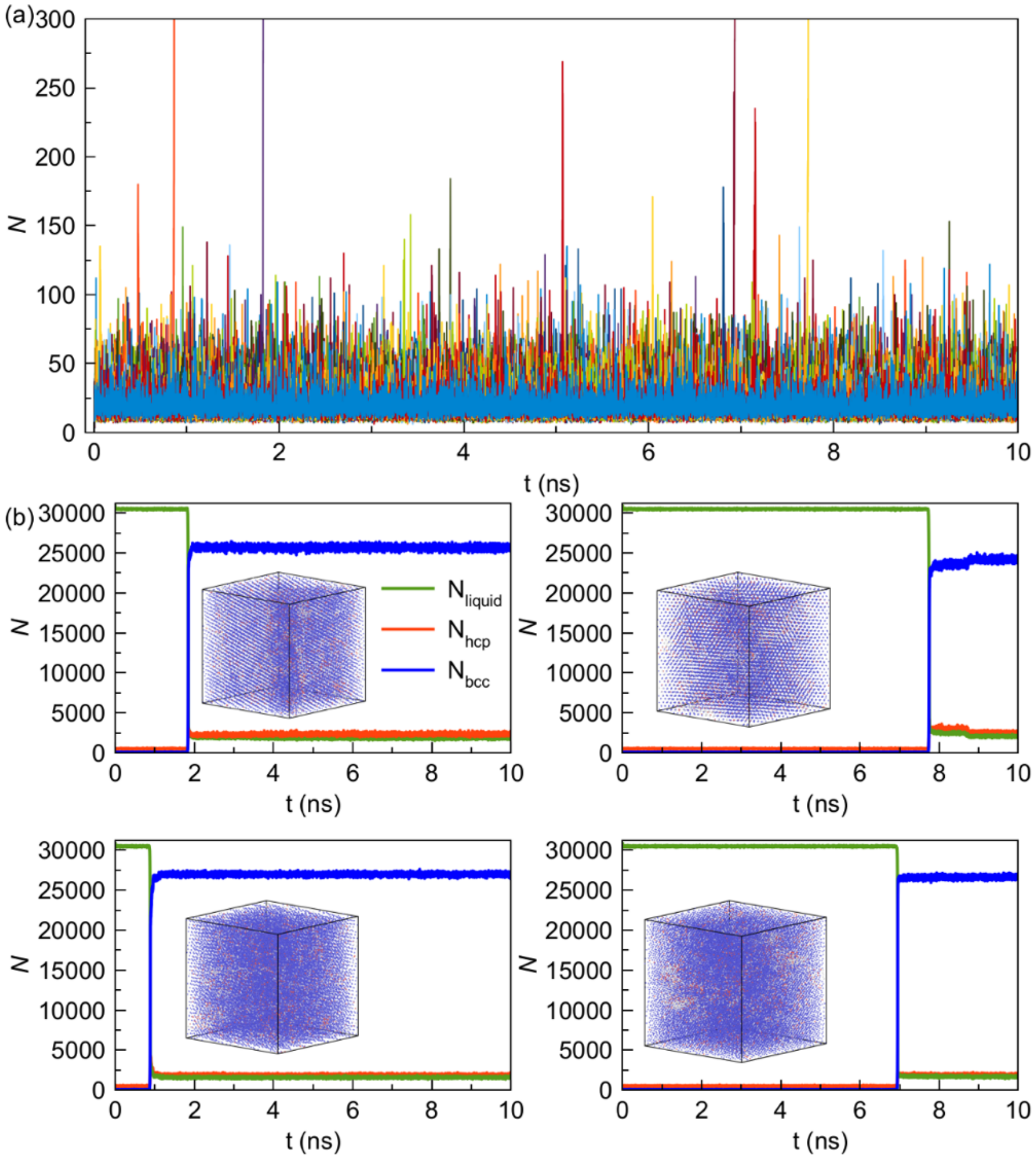

FIG. 5. (a) The nucleus size as a function of time in brute-force simulations of PotA at 4450 K. Each colored line represents an independent MD simulation. (b) The population of liquid, hcp, and bcc-like atoms as a function of time during crystallization in four nucleated trajectories.

difference between PotM and PotA, and this difference decreases with decreasing temperature.

To confirm the PEM results that bcc nucleation is much faster than hcp for PotA, we performed 60 independent brute-force simulations for PotA at a temperature of 4450 K (i.e., at the undercooling of -1756 K). The liquid, initially free of embryos, was kept at this temperature for 10 ns. The size of the largest nucleus in the system is shown as a function of time in Fig. 5(a). In such brute-force simulations, overcoming the nucleation barrier requires a considerable amount of time. Nucleation eventually occurred in only 4 out of the 60 independent simulations. In such cases, a nucleus spontaneously formed and grew larger than the critical size, transforming the system into a bulk crystal. As shown in Fig. 5(b), these simulations formed a bcc crystal. At this temperature, the liquid system can form a critical nucleus primarily dominated by bcc, while forming hcp critical nuclei is more difficult with PotA. Thus, the brute-force simulations suggest that bcc nucleates much faster than hcp in this temperature regime. While Wilson *et al.* [21] performed simulations near similar temperatures, a detailed structural analysis of the nucleated phases was not reported. At lower temperatures, nucleation becomes faster, and the systems exhibit strong phase competition between the hcp and bcc phases that the two types of nuclei can both form in the simulation.

### C. The undercooling of Earth's core nucleation

We estimate the required undercooling for Earth's core nucleation with temperature-dependent nucleation rate $J$ from PotA and PotM. Because the critical nucleus only has half a chance to grow at the top of the nucleation barrier, the waiting time in a fixed volume can be expressed as $\tau_v = \frac{1}{2J}$ [20]. Figure 6 shows the nucleation waiting time for Earth's core as a function of undercooling. The waiting time for the hcp phase obtained via PotA shows a systematic difference of 2-5 orders of magnitude compared to that of Wilson *et al.* [21]. This difference is relatively small compared to the large change in nucleation rate due to temperature variation. Thus, the PEM simulation of the hcp phase with PotA can be considered consistent with the estimation of required undercooling for IC nucleation by Wilson *et al.* [21]. However, the bcc waiting time is much smaller than that for the hcp phase. This behavior is consistent between PotA and PotM.

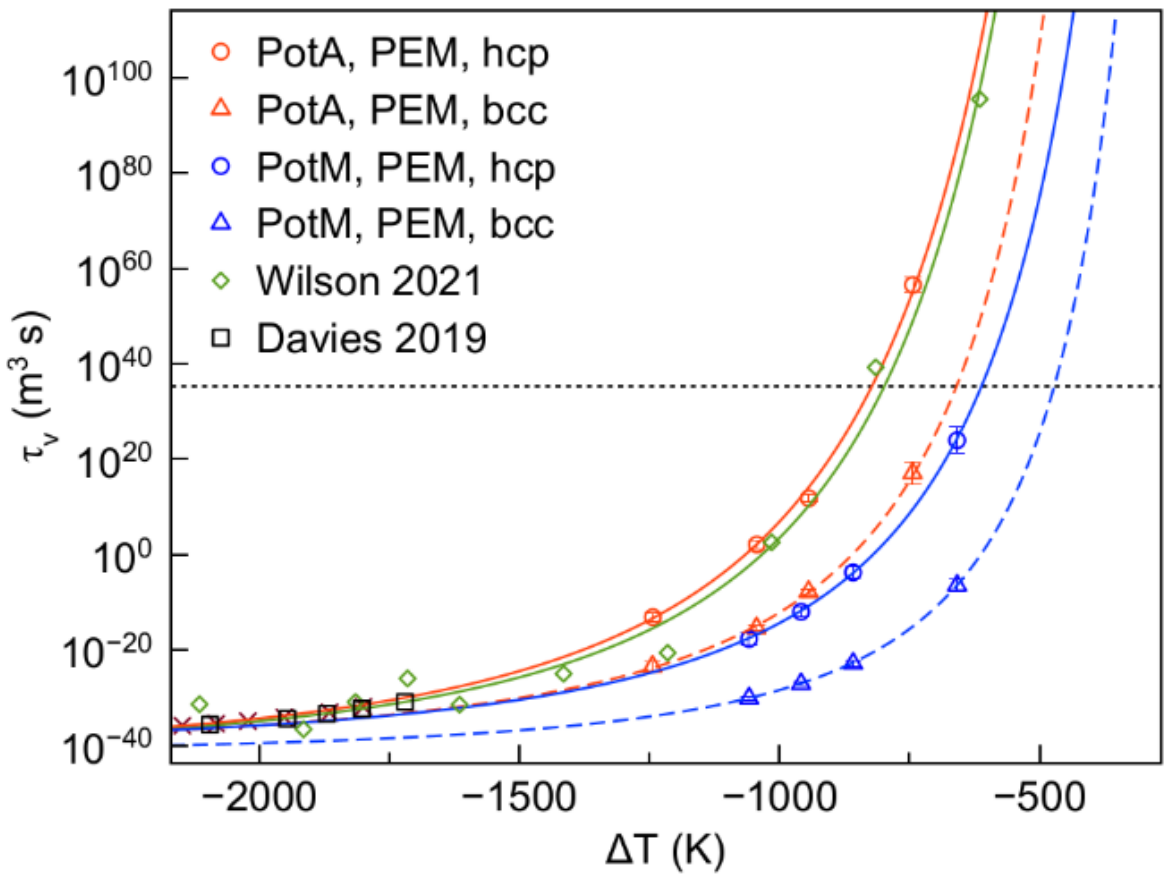

FIG. 6. The nucleation waiting time of Earth's core as a function of undercooling relative to the melting temperature of hcp. The red and blue lines are based on $\tau_v = \frac{1}{2J}$. The green markers and line are from Wilson *et al.* [21]. The black markers are from Davies *et al.* [20]. The dotted line shows the nucleation waiting time based on the assumed Earth's inner core age and volume.

The possible $\tau_v$ of the Earth's core can be estimated as follows. The nucleation incubation time is approximately one billion years, likely the upper limit of plausible inner core age according to the core's cooling rate [20]. The initial undercooled region should be smaller than the current inner core. Thus, the inner core volume of ~$7.6\times10^{18}$ m$^3$ can be used to estimate the upper limit of the nucleation volume. Therefore, $\tau_v$ of the Earth's core can be estimated to be on the order of $10^{35}$ m$^3\cdot$s, as indicated by the dotted black line in Fig. 6. The intersections of this line with the $\tau_v$ of hcp and bcc phases provide the required undercooling. With PotA, a critical undercooling of 820 K is required for the hcp phase and 660 K for the bcc phase. With PotM, a critical undercooling of 610 K is required for the hcp phase and 470 K for the bcc phase.

## IV. DISCUSSION

We have compared the key properties that determine the nucleation kinetics of Fe under Earth's IC conditions with PotA from Alfè *et al.* [20,30] and PotM from Sun *et al.* [25]. Both potentials predict a higher melting temperature for the hcp phase than the bcc phase, indicating that the hcp phase is more stable than the bcc phase against melting under IC conditions. The difference in melting points can contribute to the different nucleation rates described by the two potentials. However, even if the two potentials describe the same melting temperature, it does not guarantee that their nucleation rates will agree, as nucleation depends on the complex interplay between the temperature dependence of the chemical potential and the SLI free energy. Over a wide range of undercooling, both potentials show that the chemical potential of the bcc phase is systematically higher than that of the hcp phase, further confirming that hcp is the stable phase for pure Fe, while bcc is a metastable phase. Similar results, confirming the stability of the hcp phase over the bcc phase for pure Fe under IC conditions, have been reported in several recent studies using *ab initio* and deep-learning simulations [47–51].

The slope of the chemical potential with respect to undercooling for PotM is steeper than that for PotA, resulting in a stronger driving force of nucleation for PotM compared to PotA at the same undercooling. This is one of the reasons why the nucleation rate of PotM is systematically higher than that of PotA. For both potentials, the SLI free energy of hcp is systematically larger than that of bcc. Even though the driving force (chemical potential difference) for nucleating hcp increases at deeper undercooling, the strong energy penalty due to the SLI free energy keeps the hcp phase at a higher nucleation barrier.

Neither the bcc nor the hcp phase forms a spherical nucleus in our simulations; however, the shape approaches sphericity with increasing nucleus size, as shown in Fig. 2. The assumption of spherical nuclei in Wilson *et al.* [21] can introduce differences in nucleus size and nucleation rate calculations compared to our results for PotA. This effect can be more significant in impure systems, as illustrated in their recent work [22]. Therefore, a non-spherical correction should be applied to the CNT formula when the nucleus size is small. On the other hand, while the different methods used to compute nucleation rates here and in Wilson *et al.* [21] can cause a few orders of magnitude difference in the nucleation rate of hcp, such difference causes only a small change in the nucleation waiting time vs. temperature, as shown in Fig. 6.

Although PotA and PotM differ in their descriptions of chemical potential and SLI free energy of Fe under IC conditions, both potentials predict a faster nucleation rate and shorter nucleation waiting time for the bcc nucleus than for the hcp nucleus. This is primarily due to the smaller SLI free energy penalty, which leads to a lower nucleation barrier for the bcc phase. Therefore, the two-step nucleation mechanism of pure Fe under IC conditions [25] is consistently predicted by two potentials. Most high-pressure experiments using shock-wave and diamond anvil cells have only observed the hcp phase of Fe at high pressures [52]. Recently, XRD peaks of bcc Fe were identified in a high-pressure experiment using the pump-and-probe mode of XFEL pulses, where the sample underwent periodic heating and cooling [53]. This process creates conditions for repeated melting and recrystallization of the sample, providing an ideal way to study nucleation. The high-pressure bcc phase only appears under these specific conditions. In contrast, the hcp-to-bcc phase transition has never been observed in heating experiments under static conditions, such as direct laser-heated diamond anvil cell experiments. This combined experimental evidence suggests that the bcc phase nucleates much faster than the hcp phase but remains a metastable phase for pure Fe at high pressure, consistent with our two-step nucleation scenario. However, this does not rule out the existence of the bcc phase in the IC, as suggested by recent studies of Fe alloyed with other elements (see, e.g., [23,54]).

Considering the slow cooling rate of Earth's core, approximately 100 K $Gyr^{-1}$, the initial undercooling is expected to range from 100 K to 200 K [22]. Although the two potentials predict different critical undercooling values, both fall outside the reasonable undercooling range. This suggests that the pure Fe model cannot solve the nucleation paradox of Earth's core. Therefore, it is crucial to consider the contribution of other elements to the core nucleation process [24]. Nonetheless, the influence of the bcc phase, with its faster nucleation rate, should also be considered.

## V. CONCLUSIONS

In summary, we investigate the performance of two distinct potentials, each developed with different datasets and methods in simulating the nucleation of Fe under IC conditions. For both potentials, the melting temperature of the bcc phase is lower than that of the hcp phase. The chemical potential of the bcc phase is higher than that of the hcp phase within a large undercooling range, suggesting that the hcp phase is the stable phase while the bcc phase is a metastable phase for pure Fe at IC conditions. Both potentials predict that the bcc phase has a lower SLI free energy. These effects lead to a smaller critical nucleus size, a lower nucleation barrier, a faster nucleation rate, and a shorter nucleation waiting time of the bcc phase compared to those of the hcp phase. The results from brute-force simulations also suggest that the bcc phase nucleates faster than the hcp phase. However, the nucleation driving force of PotM is stronger than that of PotA for both bcc and hcp. The SLI free energy of the bcc phase for PotA is higher than that for PotM. Thus, PotM estimates a systematically larger nucleation rate than PotA. However, both potentials predict different undercooling values for IC nucleation by ~100 K, falling outside the reasonable undercooling range that Earth's core can reach in its cooling history. Therefore, it is crucial to consider the contribution of other elements and the possibility of two-step or even multistep nucleation in future studies.

## ACKNOWLEDGMENTS

Work at Xiamen University was supported by the National Natural Science Foundation of China (Grants Nos. T2422016 and 42374108). RMW acknowledges support from the Gordon and Betty Moore Foundation Award GBMF12801 (doi.org/10.37807/GBMF12801) and National Science Foundation (Grants Nos. EAR-2000850 and EAR-1918126). S. Fang and T. Wu from the Information and Network Center of Xiamen University are acknowledged for their help with Graphics Processing Unit computing. Calculations were performed on Bridges-2 system at PSC, the Anvil system at Purdue University, the Expanse system at SDSC, and the Delta system at NCSA through allocation TG-DMR180081 from the Advanced Cyberinfrastructure Coordination Ecosystem: Services & Support (ACCESS) program, which is supported by National Science Foundation grants #2138259, #2138286, #2138307, #2137603, and #2138296. The supercomputing time was partly supported by the Opening Project of the Joint Laboratory for Planetary Science and Supercomputing, Research Center for Planetary Science, and the National Supercomputing Center in Chengdu (Grants No. CSYYGS-QT-2024-15).